%% file: 18_kar.tex
\documentclass{ragtime}

\usepackage{amsthm}
\usepackage[mathscr]{euscript}
\usepackage{times}
\usepackage{txfonts}
\usepackage{color}
\usepackage[T1]{fontenc}
\usepackage[utf8]{inputenc}
\usepackage[polish,english]{babel}
\usepackage{subcaption}
\usepackage{caption}
\usepackage{wrapfig,booktabs,multirow}

\newcommand{\koral}{\texttt{KORAL}}
\newcommand{\msol}{{\mathrm{M}}_\odot}
\newcommand{\pgas}{{p_{\mathrm{gas}}}}
\newcommand{\uint}{{u_{\mathrm{int}}}}
\newcommand{\rg}{r_{\mathrm{g}}}
\newcommand{\tg}{t_{\mathrm{g}}}
\newcommand{\der}{\mathrm{d}}

\hypersetup{
        colorlinks=true,
        allcolors=blue,
        }

\title[GRRMHD resolution study]
{General Relativistic Magnetohydrodynamic Simulations of Accreting Tori: Resolution Study}

\author[A. Karakonstantakis, D. Lan\v{c}ov\'a, M. \v{C}emelji\'{c}]
       {Angelos Karakonstantakis\at{1,a}
       Debora Lan\v{c}ov\'a\at[]{2,b} and 
     \splitauthors
       Miljenko \v{C}emelji\'{c}\at[]{2,1,3} \splitauthors
       \\ 
        \ins{1}Nicolaus Copernicus Astronomical Centre, Polish Academy of Sciences,\splitins[1] ul. Bartycka 18, 00-716 Warsaw, Poland \\
       \ins{2}Research Centre for Computational Physics and Data Processing, \splitins[3]\hspace{0.8mm}Institute of Physics, Silesian University in Opava, 
       \splitins[2] Bezru\v{c}ovo n\'am.~13, CZ-746\,01 Opava, Czech Republic\\
        \ins{3}Academia Sinica, Institute of Astronomy and Astrophysics,\splitins[3] P.O. Box 23-141, Taipei 106, Taiwan\\
        \ins{a}\Email{karakonang@camk.edu.pl}\\
        \ins{b}\Email{debora.lancova@physics.slu.cz} } 

\coentry{A. Karakonstantakis, D. Lan\v{c}ov\'a, M. \v{C}emelji\'{c}}{General Relativistic Magnetohydrodynamic Simulations of Accreting Tori: Resolution Study}

\graphicspath{{18_kar/}}

\begin{document}

\begin{abstract}
    We present two-dimensional general relativistic radiative magnetohydrodynamical simulations of accretion disks around non-rotating stellar-mass 
     black hole. We study the evolution of an equilibrium accreting torus in different grid resolutions to determine an adequate resolution to produce a stable turbulent disk driven by magneto-rotational instability. We evaluate the quality parameter, \(Q_{\theta}\), from the ratio of MRI wavelength to the grid zone size and examine the effect of resolution in various quantitative values such as the accretion rate, magnetisation, fluxes of physical quantities and disk scale-height. We also analyse how the resolution affects the formation of plasmoids produced in the magnetic reconnection events.
\end{abstract}


\section{Introduction}\label{s:intro}
Global General Relativistic (GR) Radiative (R) Magnetohydrodynamical (MHD) simulations are an excellent tool to model the violent and turbulent environment in the vicinity of an accreting compact object. Even though this method can capture the interplay of many different physical processes governing the behaviour of magnetised plasma, it also has many limitations. 

One significant limitation is that the GRRMHD solves the flux of quantities on a fixed discrete grid, neglecting the structure inside the grid cells and approximating them as homogeneous. The solution to the related Riemann problem is always only approximate; thus, some information on the low-scale fluctuations is lost. At the same time, accretion in magnetised disks is known to be driven by low-scale turbulences caused by magneto-rotational instability \cite[MRI, ][]{Balbus1991,Balbus1998}. 

The smaller cell size leads to a smaller time step. The issue becomes even more complex in the case of the GR due to singularities in the commonly used coordinate systems and limitation by the speed of light in the case of radiative MHD. A suitable choice of computational grid parameters is fundamental in GRMHD simulations, as it must have a sufficient resolution to resolve the low-scale turbulences while considering computational demands. 

The challenge of the GRMHD method is the scalability of turbulences present in the accretion flow. This makes it difficult to achieve a fully converged solution. Convergence in an accretion disk simulation is often determined by the stability of the flow parameters (e.g., mass accretion rate, density scale height, radiative luminosity). Recently, \cite{White19} studied numerical convergence in simulations of magnetically arrested disks (MAD) and its effects in the lunch Jets. They studied the effect of 4 different resolutions. They found that turbulent structure does not appear in their lowest resolution, and MRI is suppressed. Also, \cite{Ripperda22} conducted extreme-resolution \(10^4\) grid shell resolution 3D simulations of MAD flows and analyzed the appearance of blob structures. They showed that magnetic flux bundles can escape from the event horizon through reconnection and complete a full orbit as low-density hot spots consistent with interferometric observations.

Quality parameters were derived to assess simulation convergence in studies by \cite{Hawley11,Hawley13}. They are based on the assumption that the characteristic wavelength of the MRI must be well resolved in order to maintain the MRI-induced turbulences. Although satisfying these conditions is important for obtaining reliable and physical results, other limitations and approximations of the GRMHD must still be considered.

Here, we investigate the influence of increasing resolution on simulation results. The structure of this Proceeding is as follows: in Section \ref{s:setup}, we describe the employed code \koral{} and the numerical setup of our simulation. In Section \ref{s:tools}, we describe the diagnostic tools we considered; in Section \ref{s:results}, we present the analysis of the simulation results, and in Section \ref{s:conclusions}, we comment on our finding and draw conclusions. 

\section{Numerical setup and initial condition}\label{s:setup}

\subsection{\koral{} implementation}

We use the GRRMHD code \koral{} to perform global two-dimensional (2D) axisymmetric simulations of accretion onto a $M = 10\,\msol$ non-rotating black hole (BH)\footnote{$\msol$ denoting the solar mass}.
The grid resolution is logarithmic in radius and stretches up to $r_{\mathrm{out}} = 1000\,\rg$, where $\rg \equiv G M /c^{2}$ is the gravitational radius. We considered four increasing-resolution simulation grids described in Table~\ref{tab:accRate}.

\koral{} solves the conservation equations for a fluid with rest-mass density $\rho$, 4-velocity $u^\mu$, and stress-energy tensor $T^\mu{}_\nu$, coupled with the radiation tensor ($R^\mu{}_\nu$) via the radiation 4-force density $G_\nu$. The equations are
\begin{equation}
    \nabla_\mu \left(\rho u^\mu\right) = 0,
\quad
    \nabla_\mu T^\mu{}_\nu = G_\nu,
\quad
    \nabla_\mu R^\mu{}_\nu = -G_\nu.
\end{equation}
The MHD stress-energy tensor $T_\mu{}_\nu$ is given by
\begin{equation}
     T^\mu{}_\nu = \left(\rho + \uint + \pgas + b^2\right) u^\mu u_\nu + \left(\pgas + \frac{1}{2} b^2 \right) \delta^{\mu}_{\nu} - b^\mu b_\nu,
\end{equation}

\noindent where \(b^\mu\) is the magnetic field four-vector, \(\uint\) and \(\pgas = (\gamma - 1) \uint\) are the internal energy and pressure of the gas in the comoving frame, with $\gamma$ being the adiabatic index. 

The implementation of the various processes in the \koral{} code was described in an extensive collection of published papers, including \cite{Koral13,Koral14,Sadowski2015_compton,Koral15,Koral17,Chael2017}, and recently also summarised in \cite{DLthesis}.

\subsubsection{Mean-field dynamo}

MRI-driven turbulences are responsible for angular momentum transport in BH accretion disks. A physical mechanism is required to sustain the magnetic fields in time against dissipation. Local and global simulations of accretion disks have shown that shearing due to differential rotation induces a turbulent dynamo capable of amplifying and sustaining magnetic fields against dissipation \citep[see][ and references therein]{Zanna22}.

MHD dynamo cannot operate in axisymmetric plasma configurations, as follows from Cowling's anti-dynamo theorem \citep{Brandenburg1995}. Therefore, capturing the growth of MRI self-consistently requires a three-dimensional (3D) simulation, which is computationally expensive given the long evolution times and requirements for high resolution. However, the \koral{} code implementation allows for replenishment of the magnetic field with an artificial mean-field dynamo term in the case of a 2D axisymmetric configuration, as introduced in \cite{Koral15}. The dynamo efficiently counterbalances the magnetic field dissipation. 

\subsection{Initial state}
\label{ss:init}

\begin{figure}
\centering
\includegraphics[width=0.65\linewidth]{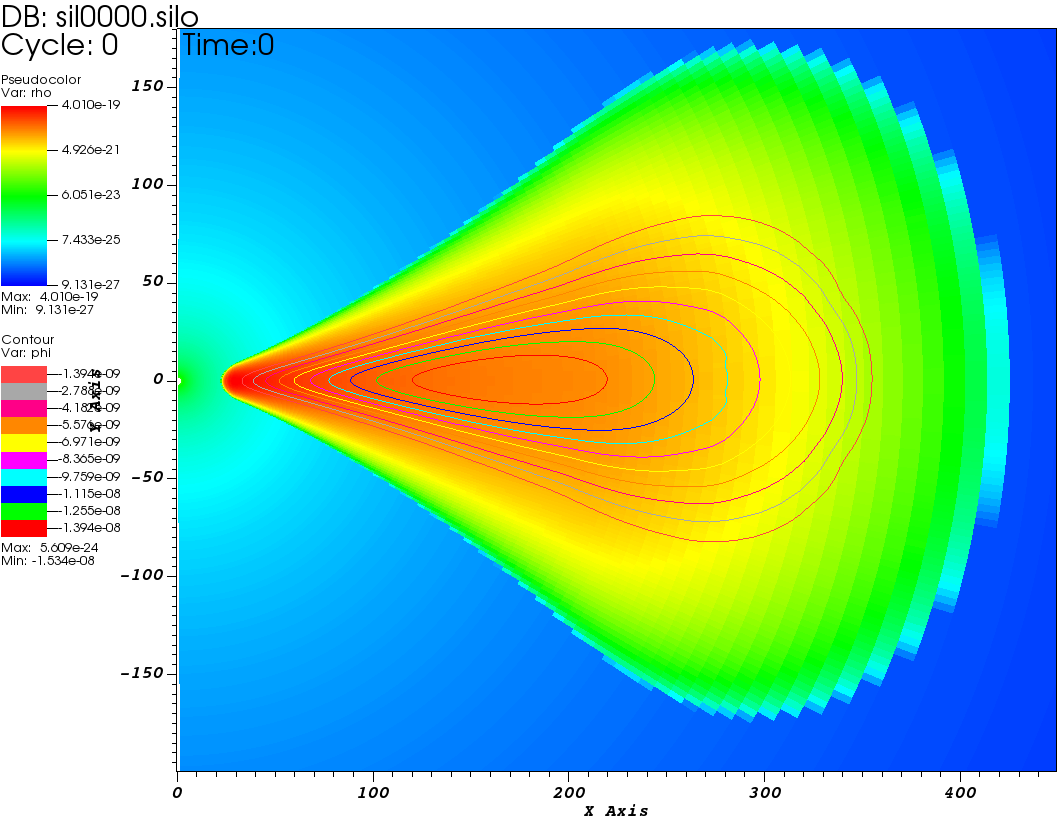}
\caption{Snapshot of the initial state in our simulations, showing logarithm of gas density. Contours show the initial poloidal magnetic field lines.}\label{fig:init}
\end{figure}

All simulations presented in this work are initialised by a torus in hydrodynamical equilibrium following \cite{Penna13}. Fig.~\ref{fig:init} shows the initial distribution of matter and magnetic field topology in the meridional plane. 
The inner edge of the initial torus is located at \(R_{\rm in} = 22\,\rg\). It is initially threaded with a poloidal magnetic field, such that the magnetization in the torus centre is $\beta= {p_\mathrm{mag}}/{p_\mathrm{gas}} = 20$, where $p_\mathrm{mag}$ is the magnetic pressure.

The initial angular momentum on the equatorial plane is set up as a fraction of the Keplerian \(\Omega\) multiplied by a factor \(\xi = 0.975\) for \(R > R_1 = 30\,\rg\). Inside \(R_1\), the angular momentum is constant. The initial state corresponds to simulation \texttt{r300a0} (but with single loop magnetic field loop as shown in Fig.~\ref{fig:init}) in \cite{Koral15}, where the results were compared to a 3D simulation with the same parameters but without the dynamo term, leading to comparable time-averaged results. The effects of different magnetic field configurations, including a similar looped configuration, have been studied by \cite{Kolos20}.

The simulations were run for {\(20\,000\)} $\tg$ in total\footnote{$\tg=\rg/c$}. After \(t \sim 10\,000\, \tg\) from the beginning, the simulations become more stable as the accretion rate stabilises to a value close to\linebreak $\dot{M} =\int_0^\pi \rho u^r \sqrt{g} \der \theta \simeq 10\, \dot{M}_{\mathrm{Edd}}$\footnote{$\dot{M}_{\mathrm{Edd}} = L_\mathrm{Edd}/ \eta c^2$,\
where,
\(\eta = - L_{\rm tot} / \int_0^\pi \rho u^r \der \theta c^2\) is the radiative efficiency with \(L_{\rm tot}\) defined in eq.~(\ref{eq:Ltot}).}. From this point to the end of the simulation, we averaged the data to study the mean properties of the accretion disk.

\section{Diagnostic tools for testing resolution}\label{s:tools}

Inadequate resolution can lead to a number of numerical artefacts, crashing of the simulations, or even wrong results. For example, it may suppress the growth of MRI when the wavelength (of at least) the fastest-growing instability mode is longer than the grid shell size. The MRI produces turbulences, but when the resolution is low, only a small range of wave numbers can be captured, distorting the properties of those turbulences. If the simulation fails to capture the turbulent structure initiated and sustained by MRI fluctuations, angular momentum is not transported outwards, which prevents mass accretion. However, high-resolution simulations are computationally very expensive. Thus, it is useful to determine whether a given simulation has adequate resolution to produce accurate quantitative results.

One would presume that numerical results converge to an exact solution as the grid size decreases. An approach to determine the accuracy of a given simulation is to run again for a higher
(e.g. doubled) resolution and check for convergence in the results. As part of this investigation, we ran a~series of simulations with increasing resolution to test the convergence. However, different quantities have different convergence rates and are subject to the limitations of the numerical algorithm used. Still, additional effects beyond resolution, such as resistivity or viscosity terms, significantly impact magnetic stress for the same resolution. Moreover, the chaotic nature of the turbulent processes introduces an additional layer of complexity in interpreting the convergence. These intricacies make it challenging to reach a definitive conclusion regarding the effect of resolution alone. 


\subsection{MRI quality parameter}\label{ss:qtheta}

The first diagnostic tool considered for resolution is derived from the MRI characteristic wavelength, \(\lambda^{\rm MRI} \propto |u_{\mathrm{A}}|/\Omega\), where \(u_{\mathrm{A}}\) is the Alfvén velocity and \(\Omega\) is the angular velocity. \cite{Sano04} studied the linear evolution of MRI {in local shearing box simulation} and found that at least 6 resolution grid zones should resolve MRI wavelength (\(\lambda^{\rm MRI}_i / \Delta x_i \geq 6\)).

Common indicators for the proper resolution to resolve the MRI turbulences are the quality parameters \(Q_i \equiv \lambda^{\rm MRI}_i / \Delta x_i\)
\citep[first introduced in][]{Nobble10,Hawley11}, where
\(\Delta x_i\) denotes the grid size in the $i$-th  direction. \cite{Nobble10} studied the properties of the turbulence through global GRMHD simulations and suggested that \(Q_{\theta} \gtrsim 20\) is adequate for resolving MRI. 

Furthermore, \cite{Hawley11}, {who performed global MHD simulations using a~pse\-udo-Newtonian potential and examined previous results of shearing box simulations}, emphasise the importance of considering a resolution parameter in the toroidal direction (i.e. \(Q_{\phi} \gtrsim 20\) and \(Q_{\theta} \gtrsim 10\)). The authors noted that the two parameters (\(Q_{\phi},\ Q_{\theta}\))  are not independent of each other: higher values in one direction can compensate for lower values in the other. For example \(Q_{\theta} \gtrsim 10-15 \) can compensate for a lower value of \(Q_{\phi} \approx 20\) \citep{Sorathia12}. Hence, the product of  \(Q_{\theta} Q_{\phi} \gtrsim 200\) has also been suggested as a criterion for convergence \citep{Narayan12}.

In our case of 2D axisymmetric simulations, only \(Q_{\theta}\) can be defined. Therefore, we consider   \(Q_{\theta}>20\) as adequate for resolving MRI. 
We define \(Q_{\theta}\) as:
\begin{equation}\label{eq:qtheta}
    Q_{\theta} = \frac{2 \pi}{\Omega \der\hat{x}^{\theta}} \frac{\left|\hat{b}^{\theta}\right|}{\sqrt{\rho}},
\end{equation}
\noindent where the magnetic field component $\hat{b}^{\theta}$ and the grid cell size $\der\hat{x}^{\theta}$ are evaluated in the fluid frame (denoted with hats).
In addition, we consider the theta average of the quality parameter as a function of radius. For this reason, we compute the density-weighted quality parameter as a function of the cylindrical radius \(R\):
\begin{equation}\label{eq:qtheta-line}
    \left< Q_\theta \right> \left(R\right) = \frac{\int Q_{\theta} \rho \der\theta}{\int \rho \der\theta}. 
\end{equation}


\section{Results}\label{s:results}

In this section, we describe selected properties of the accretion flow in the simulations and their dependency on resolution. The quantities are shown either as a function of time or their spatial or radial profile from time-averaged data.

\begin{figure}[t]
\centering
\includegraphics[width=0.85\linewidth]{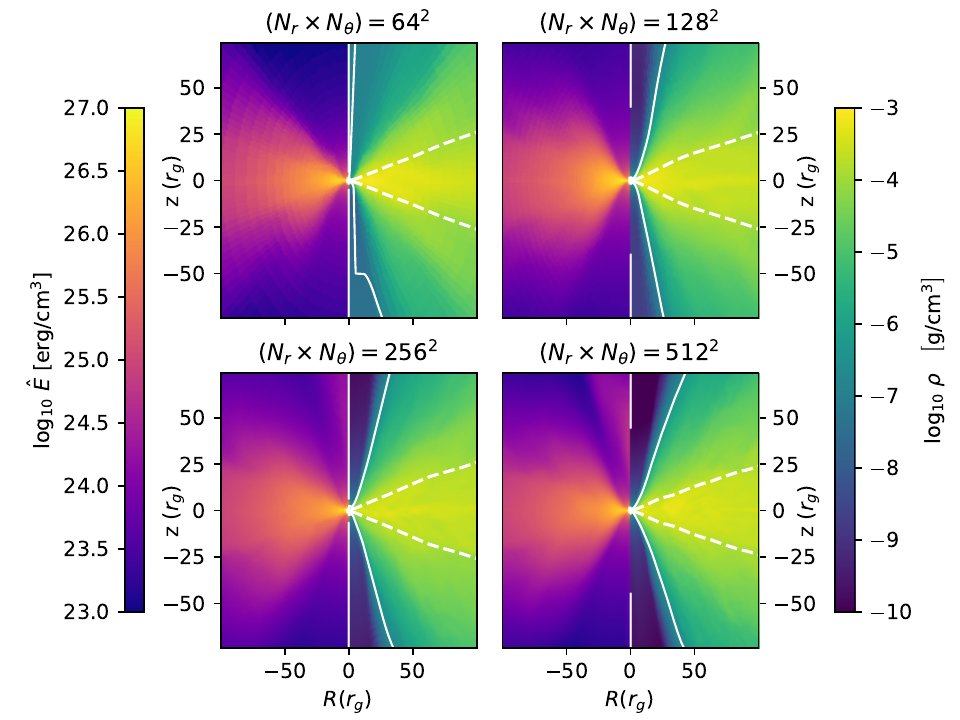}
\caption{The logarithm of radiation energy density in the fluid frame (\textit{left panels}) and time average density (\textit{right panels}). The solid line surface shows the position of the photosphere and the dashed line correspond to the density scale-height.}\label{fig:rho-ehat}
\end{figure}

Fig.~\ref{fig:rho-ehat} shows the color-maps of the gas density and the radiation energy density in the fluid frame (\(\hat{E}\)) from averaged data. The white solid line shows the surface of the last scattering (photosphere) and the dashed lines show the surface corresponds to density scale-height (\(h_\tau\)). $h_\tau$, calculated in the $z$ direction. It is apparent that for a low resolution, the funnel region is not formed properly, and the disk photosphere is much higher than in the case of the high resolution. In the {\((N_r \times N_\theta) = 64^2\)} grid shells simulation, the photosphere position cannot be properly established at all, leading to the jump in the bottom half of the domain.
We also examined the scale-height $h_{\rho}$ of the simulated accretion disk, calculated as
\begin{equation}
h_\rho =\sqrt{\frac{\int \rho z^2 \der z}{\int \rho \der z}},
\end{equation}
as shown with the dashed lines in Fig.~\ref{fig:rho-ehat} and as a function of radius in the right panel of Fig.~\ref{fig:hr} with the various solid lines corresponding to different resolutions. The time average scale-height is roughly stable for \(R > 20\, r_g\), and a typical value is between 0.25--0.30.

\begin{figure}
    \centering
    \includegraphics[width=0.85\linewidth]{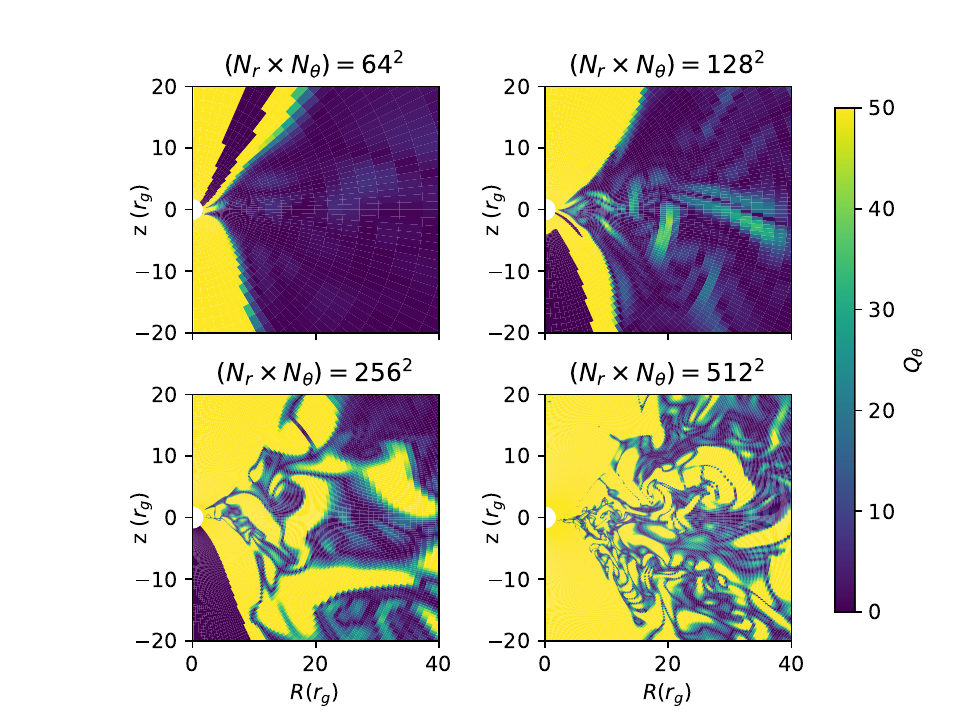}
    \caption{The values of resolution quality parameter \(Q_{\theta}\) measured at \(t = 15 \times 10^3\,\tg\).}\label{fig:qtheta-grid}
\end{figure}

Before discussing other results, we examine if the resolution is adequate for capturing MRI following the discussion in Section~\ref{s:tools}. Fig.~\ref{fig:qtheta-grid} shows the values of \(Q_{\theta}\) for each resolution. For higher resolutions, i.e. \((N_r \times N_\theta) = 256^2,\ 512^2\), the values are higher than 20 for the vast majority of the cells, which indicate adequate resolution (see Section \ref{ss:qtheta}). 

For the highest resolution, \(Q_\theta >1000\), while for the lower resolution, the values are \(<10\). We can conclude that the resolution of \(64^2\) is inadequate to capture MRI. In the case of \((N_r \times N_\theta) = 128^2\), \(Q_\theta \sim 10\) for most cells.  However, in the equatorial plane, the values reach \(Q_\theta \geq 20\). MRI may be properly resolved in that case, especially if the resolution in the toroidal direction is high enough (i.e. \(Q_\phi \gtrsim 20\)), which can compensate for lower values in \(Q_\theta\). We conclude that the resolution requirement is \((N_r \times N_\theta) = 256^2\) or greater.

\begin{figure}
\centering
    \begin{subfigure}{0.45\textwidth}
        \includegraphics[width=\linewidth]{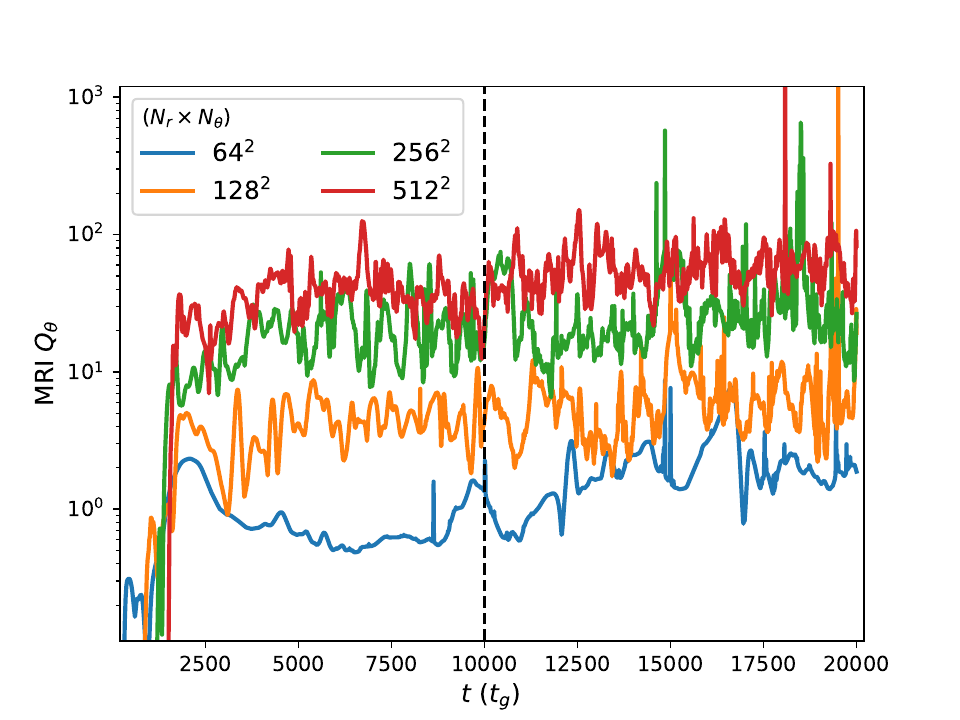}
    \end{subfigure}
    \begin{subfigure}{0.45\textwidth}
        \includegraphics[width=\linewidth]{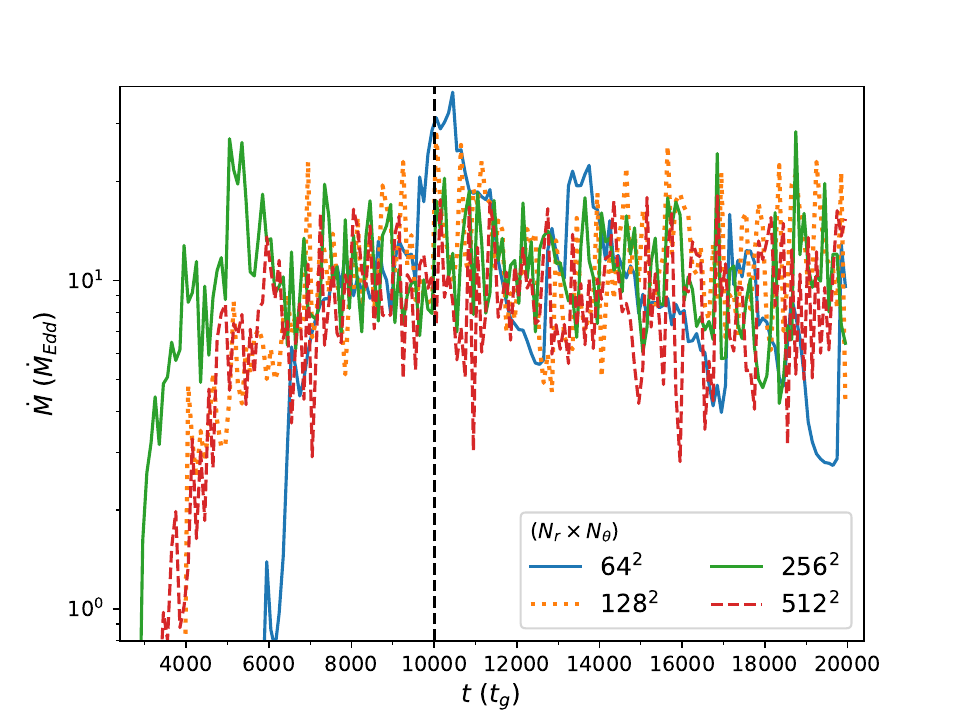}
    \end{subfigure}
    \caption{The time evolution of the quality parameter \(\langle Q_\theta \rangle (R=20\,\rg)\) (\textit{left panel}) and of the accretion rate measured close to the black hole event horizon (\textit{right panel}). The vertical line indicates\linebreak \(t \geq 10\,000\,\tg\) for which we average the results. Dashed lines were chosen for better clarity between overlapping lines.}\label{fig:qtheta-line}\label{fig:accRate-t_dotted}
\end{figure}

In Fig.~\ref{fig:qtheta-line} (left panel), we show the time evolution of the \(Q\) parameter measured at\linebreak \(R = 20\,r_g\) and theta as defined from eq.~\ref{eq:qtheta-line}.
We observe that the values are initially very low and fluctuate until the disk is formed and reaches inflow equilibrium for the measured radius, coinciding with the time when the accretion rate becomes stable.

\begin{table}
\centering
\caption{The \(10 - 20 \times 10^3\, \tg\) average and standard variance of the accretion rate at event horizon and resolution parameter at \(R= 20\,\rg\).}\label{tab:accRate}
\begin{tabular}{c c c}
\toprule
\multirow{2}{4em}{(\(N_r \times N_\theta\))} & \( \langle \dot{M} \rangle\) & {\(\langle Q_\theta \rangle\)}\\
& (\(\dot{M}_{\mathrm{Edd}}\))& \((R = 20\, \rg)\) \\\midrule
\(64^2\) & \(11.3 \pm 7.2\)& \(1.5  \pm  1.1\)\\
\(128^2\) & \(12.6 \pm 5.2\)& \(06 \pm 38\)\\
\(256^2\) & \(11.2 \pm 4.4\)& \(25 \pm 32\)\\
\(512^2\) & \(09.4 \pm 3.8\)& \(100 \pm 100\)\\\bottomrule
\end{tabular}
\end{table}

\subsection{Accretion Rate}

The time evolution of the accretion rate measured close to the BH event horizon is shown in Fig.~\ref{fig:accRate-t_dotted} (right panel). 
The mean value and variance of the accretion rate after \(10\,000\, \tg\) (indicated by the vertical line) until the end of each simulation is given in the second column of Table~\ref{tab:accRate} while the third column shows the time average value of the resolution parameter. The variance of the accretion rate fluctuations is decreasing with increasing resolution. The very high variance of the resolution parameter in the highest resolutions is because a vary high value at \(t \sim 18\,000\,\tg\) . 
However, the variability of 2D simulations using the artificial dynamo term is shown to be higher than in the 3D simulations \citep{Koral15}, so these results should be interpreted with caution.

\subsection{Magnetization}

\begin{figure}
\centering
\includegraphics[width=0.45\linewidth]{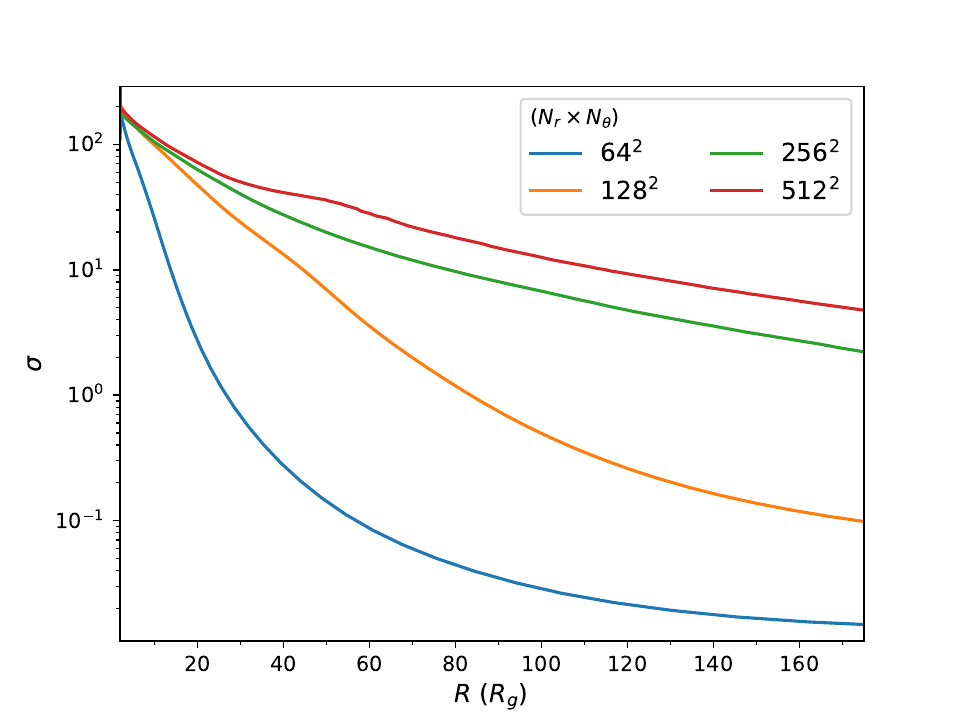}
\includegraphics[width=0.45\linewidth]{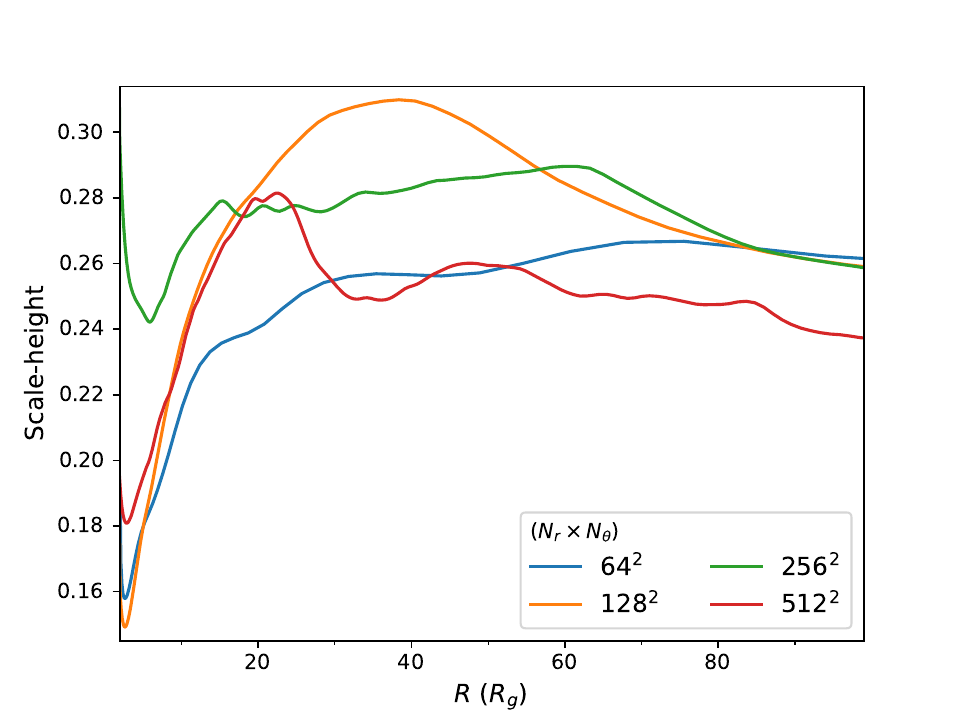}
\caption{The magnetization (\textit{left panel}) and the disk scale-height (\textit{right panel}) as a function of radius from the time-averaged data.}
\label{fig:sigma}\label{fig:hr}
\end{figure}

We calculated the magnetization, \(\sigma = b^2 / \rho\), across the resolutions. The results are plotted in Fig.~\ref{fig:sigma} (left panel). The magnetization as a function of radial coordinate decreases with increasing radius for all resolutions and the values converge as the resolution is increased with \(256^2\) and \(512^2\) grid shells resolution being almost identical. Thus, we can conclude that magnetization values are affected by resolution and tend to converge at the same value as the resolution increases.

\subsection{Luminosities}

An interesting application of numerical simulations is to compare the luminosity or spectra with observations. Therefore, in this section we examine if these results converge. The total luminosity in all forms of energy may be defined by integrating the flux carried by gas and radiation:
\begin{equation}\label{eq:Ltot}
    L_{\rm tot}\,\left(R\right) = - \int_{0}^{\pi} \left(T^r{}_t + R^r{}_t + \rho u^r\right) \sqrt{-g} \der\theta.
\end{equation}
We define the radiative luminosity:
\begin{equation}\label{eq:LR}
    L_{\rm rad}\,\left(R\right) = - \int_{0}^{\pi} R^r{}_t \sqrt{-g} \der\theta.
\end{equation}

\begin{figure}
\centering
\begin{subfigure}{0.45\textwidth}
\includegraphics[width=\linewidth]{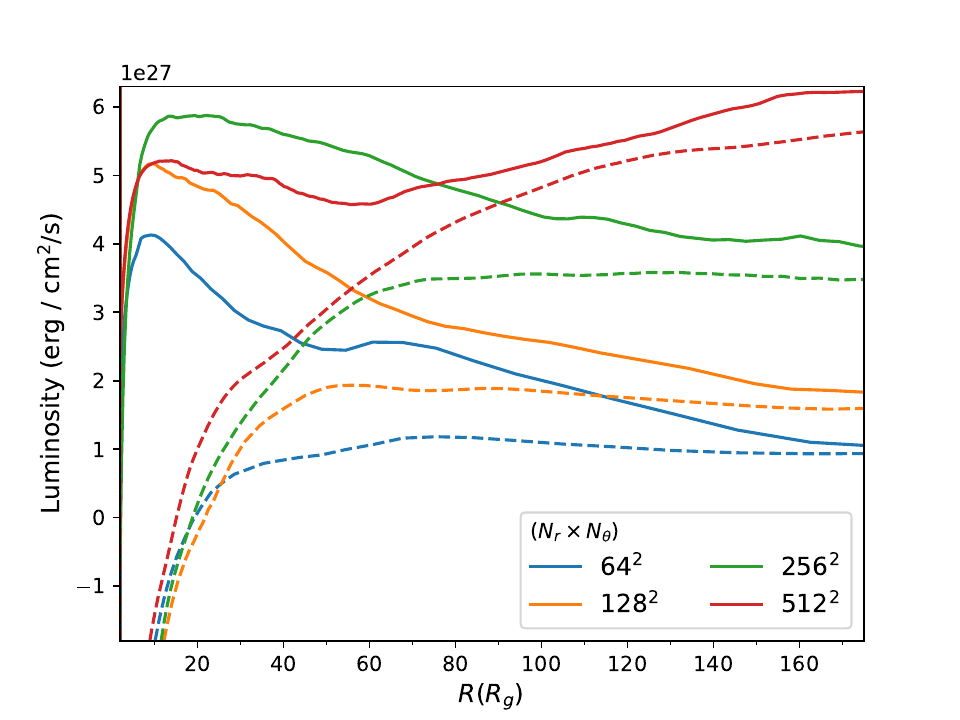}
\end{subfigure}
\begin{subfigure}{0.45\textwidth}
\includegraphics[width=\linewidth]{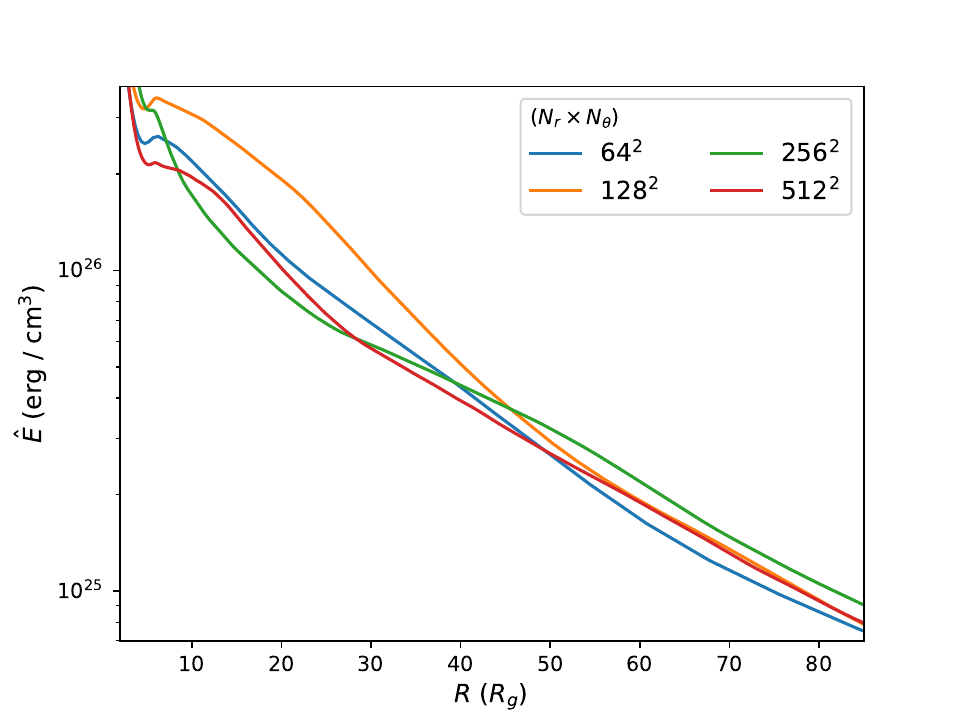}
\end{subfigure}
\caption{\textit{Left panel:} The value of total (solid lines) and radiative luminosity (dashed lines) from eqs.~\ref{eq:Ltot}, \ref{eq:LR} respectively. \textit{Right panel:} The radiation energy density in the fluid frame.}\label{fig:LR}\label{figLtot}\label{fig:ehat}
\end{figure}

Fig.~\ref{fig:LR} shows the values of \(L_{\rm tot}\) (left panel) and \(L_{\rm rad}\) (right panel) for each resolution. While these values are affected by resolution, they remain within the same order of magnitude. In the left panel of Fig.~\ref{fig:ehat}, the radiation energy density measured in the fluid frame, (\(\hat E\)), is identical for all simulations. Interestingly, the resolution does not affect the radiative quantities as much as those related to MHD fluid.  

\subsection{Formation of Plasmoids}
\begin{figure}[b]
\centering
\includegraphics[width=0.65\linewidth]{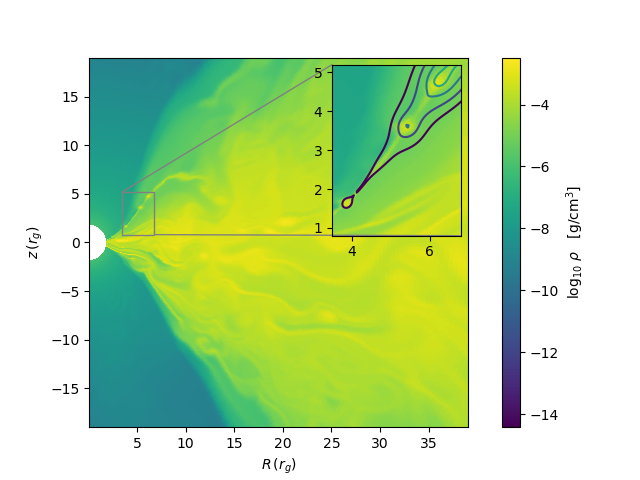}
\caption{Snapshot at \(t \sim 10\,000\,\tg\) showing the logarithm of density and the formation of plasmoids (\textit{inserted panel}) in the simulation with resolution \(N_r \times N_\theta = 512^2\) grid cells. Contours show the magnetic field lines.}\label{fig:ehat}\label{fig:plasmoids}
\end{figure}

Plasmoids, which in 2D show as magnetic islands, are bubbles of magnetised gas formed in elongated magnetic sheets, and their formation is connected with magnetic reconnection. Current sheets subject to turbulence induced in accretion disks may become thin enough to break apart into plasmoids. Because the thickness of the sheet is limited by numerical resolution, in low resolutions, plasmoids are not formed. In our results, only the highest resolution of \(512^2\) grid cells allows the current sheet to become thin enough for plasmoids to form, as shown in the right panel of Fig.~\ref{fig:plasmoids}.
This observation underscores the critical role of resolution in capturing intricate magnetic reconnection processes.

\section{Discussion and conclusions}\label{s:conclusions}

In this study, we conducted a series of radiative General Relativistic Magnetohydrodynamics (GRMHD) simulations, systematically varying the resolution to assess convergence. The selected resolutions were tested using quality parameters as resolution diagnostics, analyzed to examine the key properties of the accretion flow and determine the optimal resolution for magnetohydrodynamic simulations of the accretion disk and assessing result convergence.
We summarize our main results below:
\begin{enumerate}
    \item We observed that the accretion rate stabilizes when the Magnetorotational Instability (MRI) is properly resolved at \(t > 3 \times 10^3\, \tg\) as indicated by the value of quality parameters (\(Q_\theta > 20\)). The variance of the average value of accretion rate is decreasing with increasing resolution.
\item Resolutions lower than \((N_R \times N_\theta) < 128^2\) are inadequate to capture MRI suggesting this resolution being a potential threshold. We observed higher resolutions (i.e. \(256^2,\,512^2\)) not to have a significant quantitative difference.
\item The calculation of magnetization showed convergence as resolution increased while radiation energy density remain unaffected by resolution changes, luminosity and scale-height remained relatively stable across resolutions.
        \item Low resolution cannot resolve the fine structure of blobs, plasmoids are only appear in the highest resolution.
\end{enumerate}

This comprehensive analysis demonstrates the critical importance of resolution in accurately simulating and interpreting the dynamics of accretion flows. Adequate resolution is essential for capturing phenomena such as MRI or fine structural details such as the formation of plasmoids, ensuring the reliability of simulation results in understanding the intricate processes governing accretion disks.

\ack
This work was supported in part by the Polish NCN grants 2019/33/B/ST9/01564 and 2019/35/O/ST9/03965. M\v{C} acknowledges the Czech Science Foundation (GA\v{C}R) grant No.~21-06825X. DL acknowledge the internal grant of Silesian University, $\mathrm{SGS/31/2023}$.
A.K. thanks W{\l}odek Klu{\'z}niak and Jiří Horák for the suggestions, and useful conversations.
We gratefully acknowledge Poland’s high-performance computing infrastructure PLGrid (HPC Centers: ACK Cyfronet AGH), for providing computer facilities and support within computational grant no. $\mathrm{PLG}/2023/016168$. Part of the computations for this article have been performed using computer cluster at CAMK PAN.
\input{refdef}

\bibliography{18_kar}

\end{document}

%% file: refdef.tex

\def\prc{Phys. Rev. C }
\def\pre{Phys. Rev. E }
\def\prd{Phys. Rev. D }
\def\jcap{Journal of Cosmology and Astroparticle Physics }
\def\apss{Astrophysics and Space Science }
\def\mnras{Monthly Notices of the Royal Astronomical Society }
\def\apj{The Astrophysical Journal }
\def\aap{Astronomy and Astrophysics }
\def\actaa{Acta Astronomica }
\def\pasj{Publications of the Astronomical Society of Japan }
\def\apjl{Astrophysical Journal Letters }
\def\pasa{Publications Astronomical Society of Australia }
\def\nat{Nature }
\def\physrep{Physics Reports }
\def\araa{Annual Review of Astronomy and Astrophysics}
\def\apjs{The Astrophysical Journal Supplement}
\def\na{New Astronomy}

\def\mdash{---}